\documentclass[aps,prl,twocolumn,showpacs,superscriptaddress]{revtex4-1}

\pdfoutput=1
\usepackage[utf8]{inputenc} 			
\usepackage[T1]{fontenc}				
\usepackage{lmodern}					
\usepackage{amsmath,amssymb}
\usepackage{graphicx}
\usepackage{float}
\usepackage[usenames,dvipsnames]{xcolor}
\usepackage{tikz}
\usepackage{hyperref}

\renewcommand\vec[1]{{\bf #1}}

\begin{document}

\title{Optimal paths on the road network as directed polymers}

\author{A. P. Solon}
\affiliation{Department of Physics, Massachusetts Institute of Technology, Cambridge, MA 02139, USA}
\author{G. Bunin}
\affiliation{Department of Physics, Technion, Haifa, 32000, Israel}
\author{S. Chu}
\affiliation{Department of Physics, Massachusetts Institute of Technology, Cambridge, MA 02139, USA}
\author{M. Kardar}
\affiliation{Department of Physics, Massachusetts Institute of Technology, Cambridge, MA 02139, USA}

\date{\today}
 
\begin{abstract}
We analyze the statistics of the shortest and fastest paths on the  road network between randomly sampled end points. 
To a good approximation, these optimal paths are found to be {\it directed} in that their lengths (at large scales) are linearly
proportional to the absolute distance between them. 
This motivates comparisons to universal features of {\it directed polymers in  random media}.
There are similarities in scalings of fluctuations in length/time and transverse wanderings, 
but also important distinctions in the scaling exponents, 
likely due to long-range correlations in geographic and man-made features.
At short scales the optimal paths are not directed due to circuitous excursions governed by a fat-tailed (power-law) probability distribution. 
\end{abstract}

\maketitle


Complex networks of nodes and links characterize a wide array of systems, 
ranging form biological examples such as neural nets of neurons and synapses in the brain or chemical reactions inside a cell, 
to social or transportation networks and the World Wide Web. 
Their connectivity in the abstract space of edges and vertices has been much studied, illuminating characteristics such
as `small-world' separation, scale-free connectivity and a high degree of clustering, which can be captured by simple
models~\cite{Watts98,Barabasi99,Albert02,Ravasz03,Newman03}. 
Comparatively, less is understood about the spatial organization of complex networks
embedded in a Euclidean space, a very active area of research~\cite{BarYam1,BarYam2} 
(see also Ref.\cite{Barthelemy11} for a review). The effect of geometry is
especially relevant when the network is strongly constrained by the
environment or when the ``cost'' to maintain edges increases
significantly with their length ({\it e.g.} rivers~\cite{Dodds00},
railways~\cite{Sen03} or vascular networks~\cite{Hunt16}). The spatial
structure of streets is another example that has been particularly
studied to gain insight into organization and development of cities~\cite{Cardillo06,Lammer06,Barthelemy08}.

Much relevant information about the shape of a network is obtained by
studying the shortest paths between its nodes~\cite{Newman03}. 
More generally, it is of practical importance to find and characterize
paths that optimize a given cost function of interest. For example, in
transportation networks, one may like to understand  properties
of paths that minimize travel time, distance, or the monetary cost to go between two points. 
An obvious application is in the development of efficient GPS routing algorithms that use
prior information on optimal paths to perform better~\cite{Geisberger08}. 
This problem is challenging as the optimal paths on  transportation networks  strongly depend
on network connectivities shaped by various factors, from natural obstacles to historical development or
differences in policy.

\begin{figure}
  \includegraphics[width=.5\columnwidth]{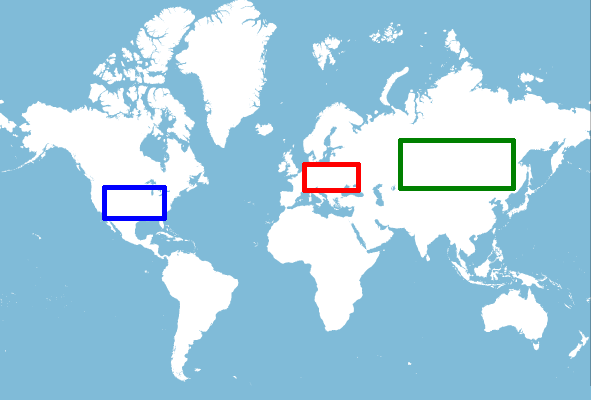}
  \caption{The location of the three regions considered. For
    simplicity and efficiency of our algorithm, they are chosen to be
    large rectangular areas (in latitude-longitude coordinates)
    without sea or ocean.}
  \label{fig:regions}
\end{figure}

The model of directed polymers in random media
(DPRM)~\cite{Kardar87,HH95} explores a physically distinct but
mathematically related problem.  It is concerned with the statistics
of a chain stretched between two points that minimizes its energy in a
random environment characterized by a rough energy landscape.  The
optimal chain configuration is then governed by a trade-off between a
line tension (preferring the straight polymer) and the hills and
valleys of the random landscape which encourage its wanderings.  A
wealth of theoretical results is available for this widely studied
problem, which belongs to the Kardar-Parisi-Zhang (KPZ)~\cite{KPZ}
universality class (originally posed to describe roughening of growing
surfaces).

Configurations of DPRM paths bear superficial resemblance to myriad natural transportation systems, 
from deltas of rivers to vascular networks; the wealth of data on road networks provides the opportunity for a quantitative comparison.
Here, we study the statistics of optimal (shortest and fastest) paths on the road network in light of known statistics for DPRM.
Gathering large data sets of millions of paths on three continents, we compute the probability
distribution of path length and travel time as a function of the (straight-line) distance between the end points. 
As a preliminary step, we confirm that long optimal paths are directed in the sense that the average length/time of
the path is (at large separations) linearly proportional to the straight-line distance.
We next examine the fluctuations of the optimized quantity.
As in the case of DPRM, appropriately scaled fluctuations can be collapsed (approximately) to a single curve,
suggesting that details of the local structure of road networks are irrelevant to the statistics on larger scales. 
However, the scaling forms do not correspond to the simplest variant of DPRM with uncorrelated energies,
indicating long-range correlations on the scale of hundred of kilometers. The transverse wanderings of the
paths is also consistent with this picture.
Remarkably, the distributions at short-scales are broad with a power-law tail of universal exponent 
that captures the profusion of long non-directed circuitous routes at separations of a few kilometers.

\begin{figure}
  \includegraphics[width=.43\columnwidth]{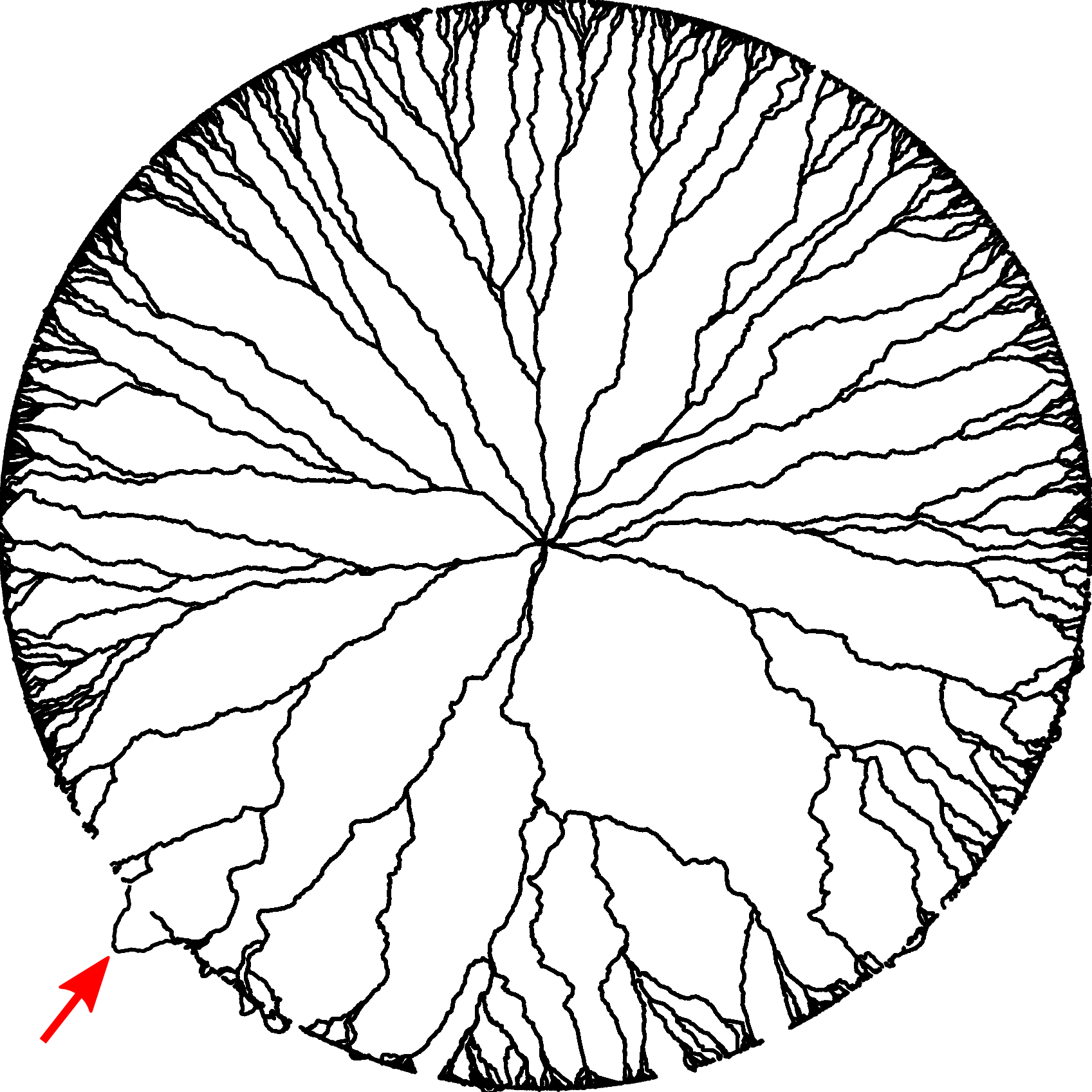}\hspace{0.5cm}
  \includegraphics[width=.43\columnwidth]{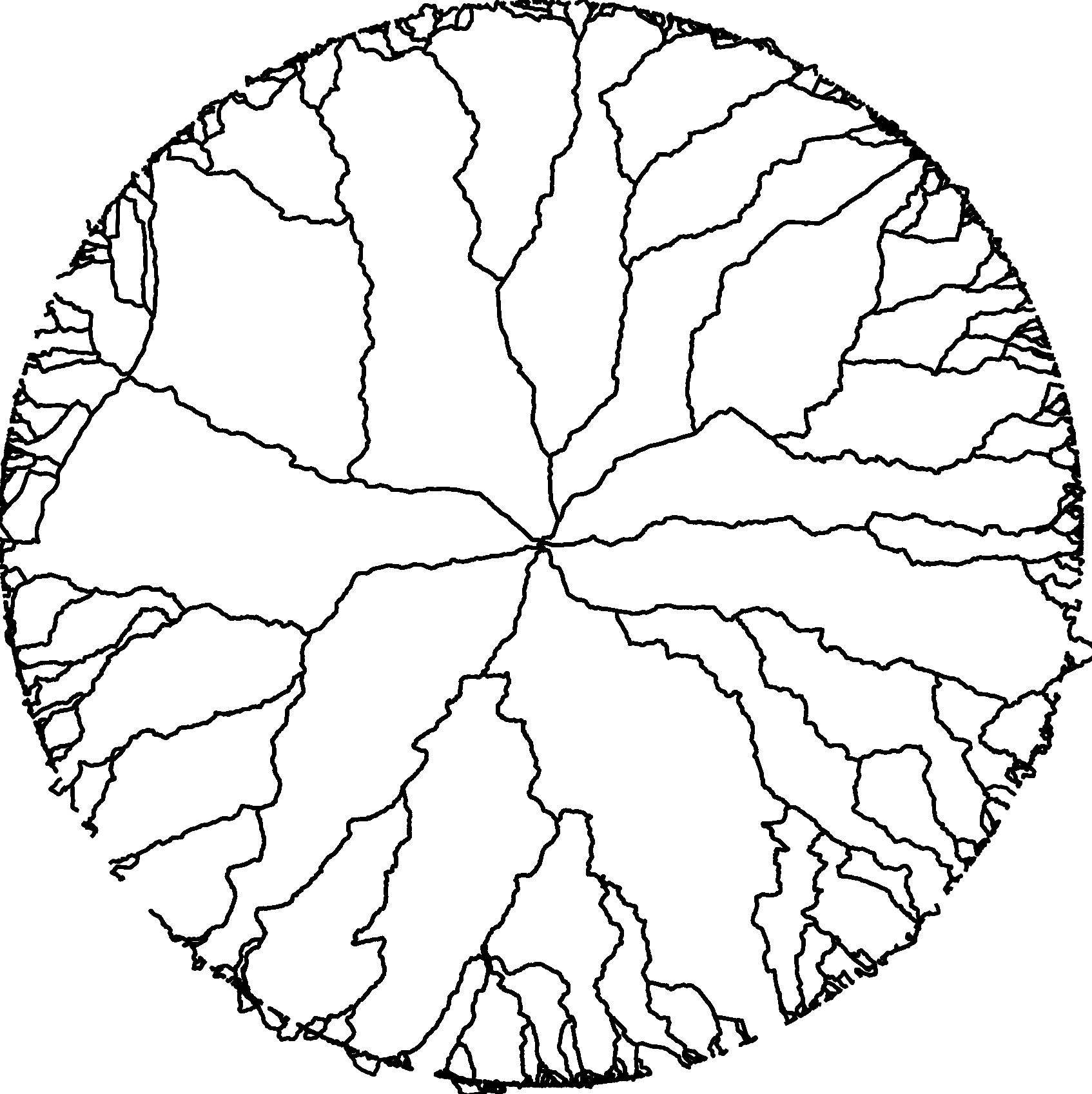}
  \caption{Shortest (left) and fastest (right) paths from a central
    point (near Munich, Germany) to $10^4$ randomly chosen points at a
    distance of $300$~km. The red arrow points to the most prominent
    overhang in the paths.}
  \label{fig:circle_path}
\end{figure}


Let us first introduce more precisely the 2-dimensional DPRM problem
and summarize the relevant results: A directed polymer is
a chain pinned at its ends, and sufficiently stretched to prevent overhangs.
Its wanderings can thus be described by a function $h(x)$, where $x$ is a coordinate along the axis between
the end points and $h$ the distance from this axis. The cost (energy) of a
configuration of  is then given by
\begin{equation}
  \label{eq:energy-DP}
  E[h(x)]=\int_0^d dx\left[\frac{\gamma}{2}\left(\frac{dh}{dx}\right)^2+V(x,h) \right]~,
\end{equation}
where $d$ is the Euclidean distance between the end points, $\gamma$ is 
the line tension of the chain and $V$ is a potential modeling a disordered environment. 
The canonical free energy at a temperature $T$ satisfies the KPZ equation~\cite{KPZ}
that  describes its scale invariant properties~\cite{foot1}.
In the zero-temperature limit, relevant to our problem, the free energy is 
simply the energy of the optimal path $E[h^*]$. 
Two exponents govern the scalings of the energy fluctuations
$\left\langle (E-\langle E\rangle)^2\right\rangle \sim d^{2\beta}$
(where the brackets denote an average over realizations of the
disorder $V$), and the transverse wanderings of the optimal chains,
$\langle h^*(x)^2\rangle \sim x^{2\zeta}$~\cite{Kardar87}.
In the simplest case of uncorrelated random potentials, the exponents
are related by the identify $\beta=2\zeta-1$, and (in 2-dimensions) given by
$\beta=1/3$ and $\zeta=2/3$~\cite{FNS77}.
More recently, it has been shown that the full distribution of $E$ is  universal,
converging at large $d$ to the Tracy-Widom (TW) distributions of random
matrix theory~\cite{Prahofer00,Takeuchi11,THH15}.
The  KPZ exponent identity can break down for correlated potentials~\cite{Medina89};
long-range correlations in $V$ lead to larger scaling exponents and different
energy distributions~\cite{Peng91,Schorr03,Kloss14,Chu16}.

\begin{figure}
  \includegraphics[width=.48\columnwidth]{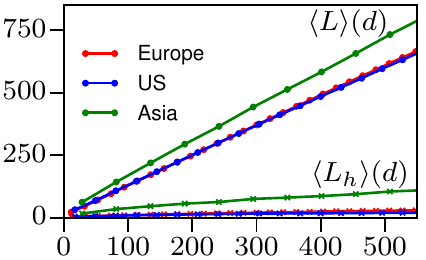}
  \includegraphics[width=.48\columnwidth]{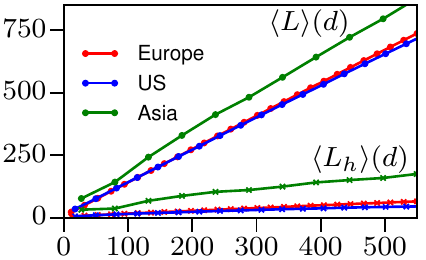}
  \caption{Average length of the optimal paths $\langle L \rangle$
    and length of overhangs $\langle L_h \rangle$ as a function of
    the distance $d$ between the end points. All lengths are measured
    in km. Left: Shortest paths. Right: Fastest paths; $10^6$ points
    for each curve. Lines are guides to the eye.}
  \label{fig:length_ovh}
\end{figure}


In light of these theoretical results, we now analyze the statistics
of two types of optimal paths (the shortest and the fastest) on the
road network. We compute the paths using the Open Source Routing
Machine (OSRM)~\cite{OSRM} operating on OpenStreetMap data, a
collaborative effort to provide an open-source map of the world. The
fastest paths are determined using the default configuration of OSRM
which takes into account speed limitations for cars and road types, but
no information on traffic. We gather six data sets for the two types
of optimal paths in the three regions indicated in
Fig.~\ref{fig:regions}, sampling the end points of the paths uniformly
on the network.

Figure~\ref{fig:circle_path} shows optimal paths from an arbitrary central point (near Munich, Germany) 
to uniformly sampled points at a distance of $300$~km. 
Both sets of optimal paths display fractal branching patterns resembling those of DPRM~\cite{Kardar87}. 
However, these
routes are not perfectly directed. This is especially visible near the
end points where the local structure of the road network may impose
overhangs (the most prominent is indicated by a red arrow in
Fig.~\ref{fig:circle_path}). Nevertheless, overhangs make a negligible contribution to
the overall optimal trajectory.
This is  quantified in Fig.~\ref{fig:length_ovh} where we plot the average
length of the paths $\langle L\rangle (d)$ and the part
$\langle L_h\rangle (d)$ corresponding to overhangs (see the
Supplementary Information for a precise definition). 
The former increases linearly with separation $d$, while the latter grows sublinearly.
Overhangs thus become less relevant at larger distances where we may expect a better correspondence between road paths and DPRM.
In the following, we divide  our study between short paths that are strongly constrained by local connectivity, 
and longer optimal paths that are directed.


\begin{figure}
  \includegraphics[width=.9\columnwidth]{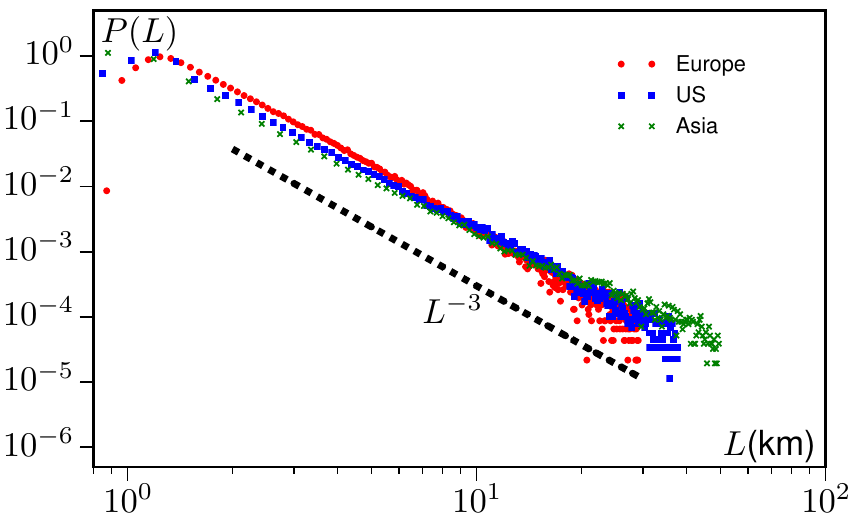}\\
  \includegraphics[width=.9\columnwidth]{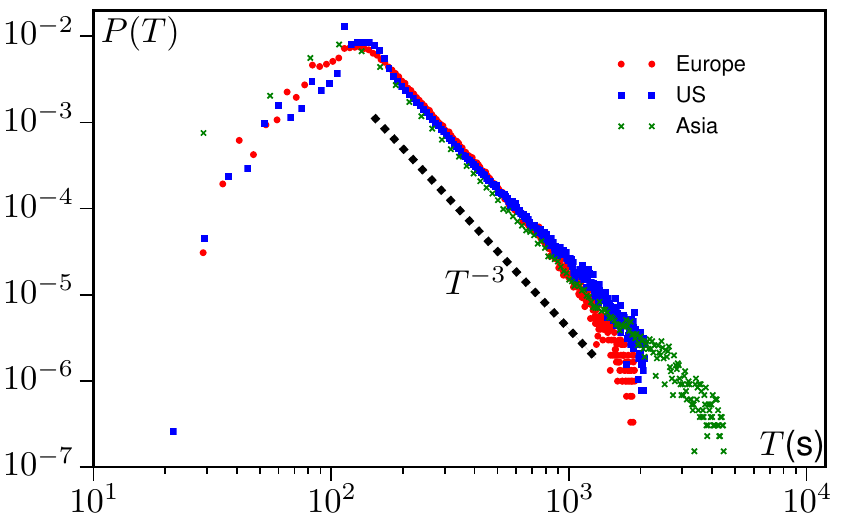}
  \caption{Optimal paths between points at a distance $d=1$ km. {\bf
      Top:} Probability distribution of the length $L$ of the
    shortest paths. {\bf Bottom:} Probability distribution of the
    travel time $T$ of the fastest paths; $N=5\times 10^5$ paths for
    each curve.}
  \label{fig:distribs_1km}
\end{figure}

We first look in Fig.~\ref{fig:distribs_1km} at the distribution of
the length $L$ of the shortest paths (or the travel time $T$ on the
fastest paths) between points at a small separation $d=1$~km. 
The distributions display clear power-law tails at large $L$
and $T$ over more than three orders of magnitude. 
The tails correspond to cases where the connecting path has to go around an obstacle to reach a
nearby point, {\it e.g.} the next bridge across a river. They
thus characterize the overhangs described previously. Most remarkably,
the decay exponent $P(L)\sim L^{-\alpha}$ (and $P(T)\sim T^{-\alpha}$)
seems to be universal across continents with $\alpha\approx 3$ (the
best fit coefficients for the six curves are all  within the range
$[2.89;3.10]$). This is surprising since we expect  paths at
small $d$ to reflect the local structure of the road network which is
{\it a priori} very different in the three regions
considered. While we lack an explanation for the value of the
exponent, we note its similarity to other problems in statistical physics: For self-avoiding random
walks, the probability of loop lengths within a long chain is governed by
$\alpha=2.68$~\cite{Duplantier87,Metzler02} while the shortest path between
nearby points on the backbone of a percolation cluster also has 
$\alpha\approx 3$~\cite{Porto98} at small distances.


Due to the fat tails in the distributions of Fig.~\ref{fig:distribs_1km}, the variances 
$\langle L^2\rangle_c(d)$ and $\langle T^2\rangle_c (d)$ are not defined,
and cannot be used to estimate an exponent $\beta$ for cost fluctuations. 
Instead, we examine the full probability distributions $P(L|d)$ and $P(T|d)$ for increasing distance $d$,
and attempt to make a ``collapse'' by superposing their maxima, and rescaling differences from the maximum by 
a factor $d^\beta$ chosen to best converge the distributions at large $d$. 
The results are shown in
Fig.~\ref{fig:distribs_convergence} (top) for the shortest paths in
Europe, and in the Supplementary Information for the five other data
sets, which show similar behavior. We find that the exponent $\beta$
can be adjusted such that the left tail of the distribution converges
rapidly to a limit distribution well-fitted by the Tracy-Widom (TW)
distribution expected for DPRM. By contrast, the
right tail converges slowly and remains heavy at the largest $d$
attainable (larger $d$, comparable to the total size of the region,
show strong finite-size effects). It is thus not clear if the right
tail also converges to TW behavior or to a different distribution, as
 observed numerically for DPRM on a long-ranged
correlated landscape~\cite{Chu16}.

\begin{figure}
  \includegraphics[width=.95\columnwidth]{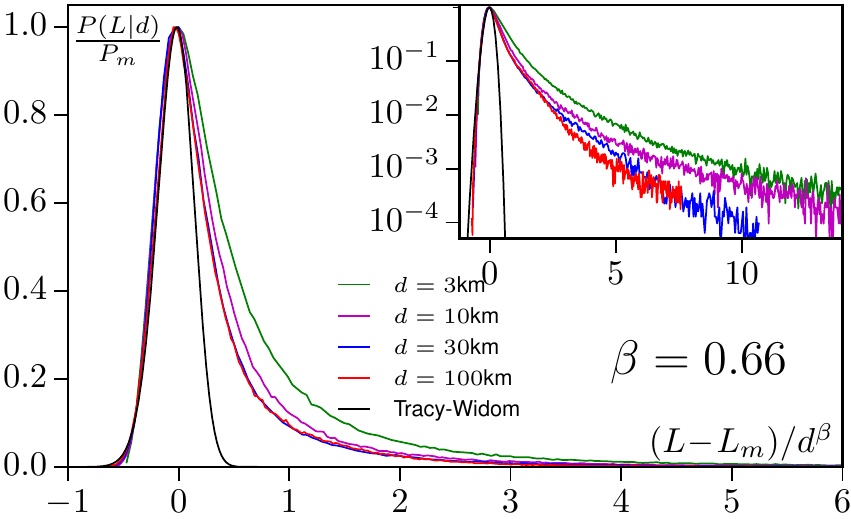}\\
  \includegraphics[width=.95\columnwidth]{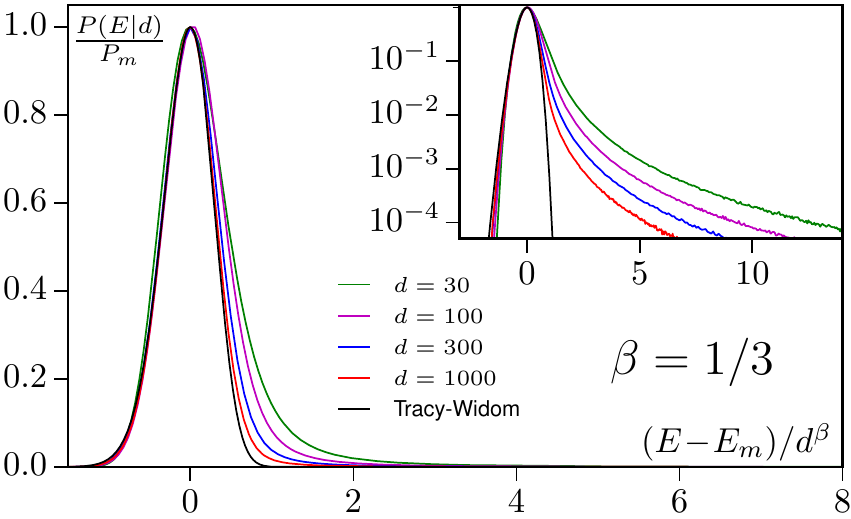}
  \caption{{\bf Top:} Probability distribution of the length $L$ of
    the shortest paths in Europe rescaled with $\beta=0.66$; $5\times
    10^5$ paths for each curve.  {\bf Bottom:} Probability
    distribution of the energy for a DPRM model (periodic boundary conditions
    on lattice of transverse size $10^7$) with power-law noise
    rescaled with $\beta=1/3$; $10^7$ paths for each curve. The insets show the same data with a
    logarithmic $y$-axis. $P_m$ denotes the maximum of the
    distribution, found at  $L=L_m$ or $E=E_m$.}
  \label{fig:distribs_convergence}
\end{figure}


For comparison, we simulated a well-established DPRM model on a square
lattice with paths  directed along the diagonal~\cite{Kardar87,Kim91}.
The distance from the diagonal (parametrized by $x$) is again denoted by $h$.
The energy of the optimal path is computed recursively as
\begin{equation}
  \label{eq:DPRM}
  E(x,h)=\text{min} \left\{ E(x-1,h),E(x-1,h-1) \right\}+\eta(x,h).
\end{equation}
After $d$ iterations, $E(d,h)$ is then the energy of the optimal path
between the point $(d,h)$ and the line $d=0$. 
To mimic the short-scale distributions in Fig.~\ref{fig:distribs_1km}, we draw the noise $\eta(d,h)$
from a power-law distribution $P(\eta)=2\eta^{-3}$ with $\eta\in [1;\infty[$. 
We then analyze the results as in the case of roads by shifting the energy distributions
$P(E|d)$ to superimpose their maxima and rescaling their width by
$d^\beta$ (Fig.~\ref{fig:distribs_convergence}, bottom).  We observe
that, as with Gaussian noise~\cite{Kim91}, the distribution converges
to a TW distribution with the KPZ exponent $\beta=1/3$. Indeed, only a
fat tail in the noise at {\it negative} energy,
$P(\eta)\sim \eta^{-a}$ as $\eta\to-\infty$ is expected to change the
scaling exponents~\cite{Pang93,Gueudre15}. Interestingly, the
convergence upon increasing $d$ is similar in the
model and the road data, with the right tails converging much
slower. This also lends credence to our measure of $\beta$ as the
exponent rescaling the left tail of the distributions.

\begin{figure}
  \includegraphics[width=.9\columnwidth]{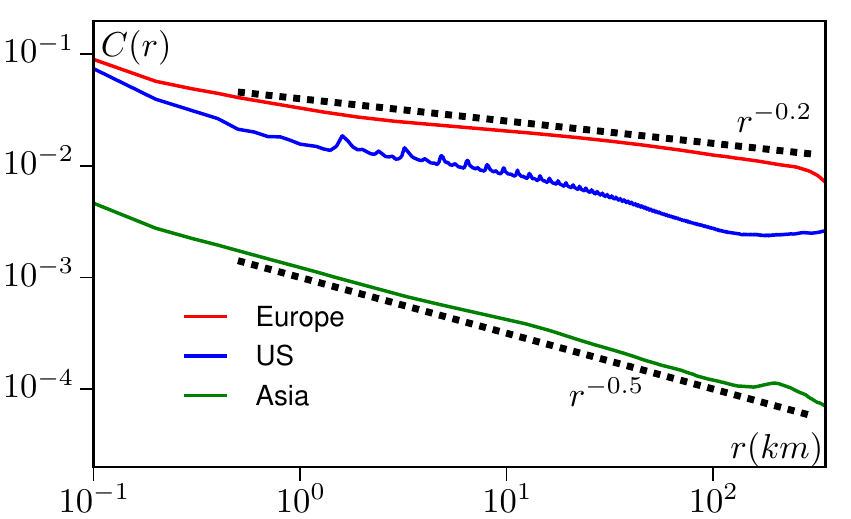}
  \caption{Auto-correlation functions of the road density as defined in
    the text. The oscillations in the curve for the US are not an
    artifact. The peaks are located every mile (with sub-peaks at
    half-miles) and correspond to large regions (up to 60 miles) of
    grid-like road network.}
  \label{fig:density_correlation}
\end{figure}

The salient distinction between the paths on the road and DPRM is the
value of the exponent $\beta$: The measured $\beta$ exponents are in
the range $0.58$ and $0.9$ (with an estimated error of $15$\%) are
much larger than $\beta=1/3$ in the (uncorrelated) KPZ universality
class.  We argue that this can be explained by the presence of
long-range correlations in the road network.  To show this, we first
discretize the full map of each region in squares of size
$100\text{m}\times 100\text{m}$ and assign the value $\rho(\vec r)=1$
if a road is found inside the square and $0$ otherwise. We then
compute the correlation function
$C(r)=\langle \rho(\vec r)\rho(\vec r+\vec x)\rangle-\langle \rho(\vec
x)\rangle^2$
where the average is taken over $\vec x$ and orientations of $\vec r$.
As shown in Fig.~\ref{fig:density_correlation}, $C(r)$ decreases
slowly (slower than $C(r)\sim r^{-0.5}$), remaining non-negligible on
the scale of hundreds of kilometers. These long-range correlations
reflect the shaping of the road network by factors acting at every
scale, from different administrative divisions to natural obstacles.
They were also shown to be important in modeling the development of
cities~\cite{Makse98,Barthelemy08}.

For DPRM, a power-law decay of correlations is known to be relevant
from both numerical simulations~\cite{Schorr03,Katsav04,Chu16} and
renormalization group analysis~\cite{Medina89,Katsav04,Kloss14}.  For
Gaussian noise with isotropic correlations decaying as a power-law
with exponents between $-0.5$ and $-0.2$ (as measured for the road
density correlations in Fig.~\ref{fig:density_correlation}), $\beta$
was measured between $0.5$ and $0.7$~\cite{Schorr03}. Given numerical
and systematic uncertainties, these values are in relatively good
agreement with our measurements for the road network.  Long-range
correlations are thus likely to be the cause of the observed large
exponents.

\begin{figure}
  \includegraphics[width=.9\columnwidth]{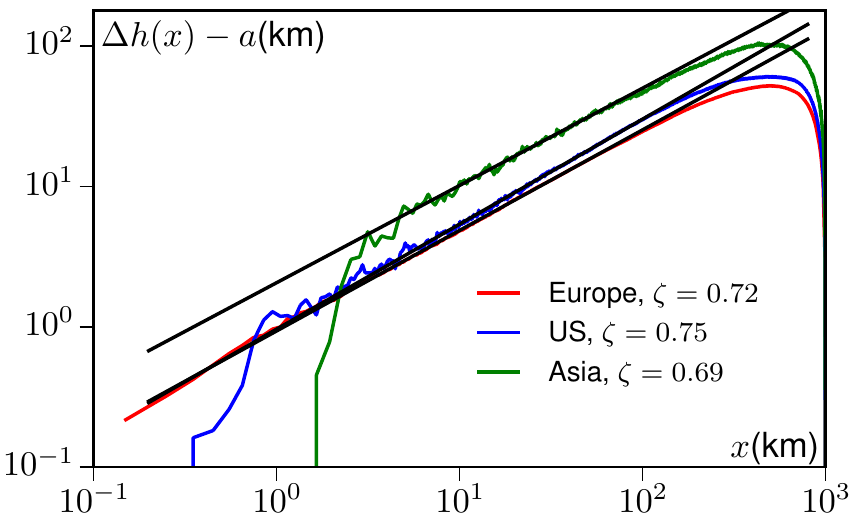}
  \caption{Transverse wanderings for the shortest paths as a function
    of the coordinate $x$ on the axis between the end points. Averages are
    over $5\times 10^5$ paths between points at distance $d=1000$~km.}
  \label{fig:height_t}
\end{figure}

Finally, we look at the wanderings of the optimal paths in the
transverse direction. The routing algorithm returns a list of points
along each path (on average every $50$m) that we use to construct the
function $h(x)$.
We do so by discretizing the distance $x$ along the end-to-end axis in bins of size $dx=100$m, and
averaging points falling in the same bin. This discards any overhangs
and  produces a directed path approximating the real path. 
For DPRM, transverse fluctuations scale as
$\Delta h(x)=\sqrt{\langle h^2(x)\rangle}\sim x^\zeta$, but
because of overhangs near the end points of road-paths $h(0)=\Delta h(0)\neq 0$ 
introducing large corrections to the putative scaling. 
Thus, as a first approximation, we estimate the exponent $\zeta$
by fitting $\Delta h(x)=a+bx^\zeta$ with free parameters $a$, $b$ and $\zeta$. 
The resulting functions $\Delta h(x)-a$ show scaling
behavior over two orders of magnitude with exponents
$\zeta\in[0.69;0.72]$ (see Fig.~\ref{fig:height_t}). Once again these
values are larger than the uncorrelated KPZ value $\zeta=2/3$, but in qualitative
agreement with the expected increase due to the  presence of long-range correlations. 
For comparison, isotropic long-range correlations with decay exponent in the range of
Fig.~\ref{fig:density_correlation} give
$\zeta\in [0.75;0.85]$~\cite{Schorr03} while correlations only in the
transverse direction yield $\zeta\in [0.67;0.72]$~\cite{Chu16}.

To conclude, we have shown that optimal paths on the road network can
be modeled as directed polymers in a random medium. To do so, we
replaced the complex road structure by a random cost landscape featuring
only the relevant properties needed to account for the observed statistics of
optimal paths. We find two important such characteristics: At short
scales, local connectivities result in circuitous paths with a scale-free
distribution of lengths characterized by a universal power-law decay, a
remarkable experimental fact that remains to be explained. This is
mimicked by a power-law distributed noise in our DPRM model. At
larger scales, the local structure becomes less relevant. The scaling
of the path length/travel time and the transverse meanderings of optimal
paths are then governed by long-range correlations that we show to be
present in the network. Although these long-range correlations are
non-universal, they show similar behaviors in the different regions of
the world considered in this paper, leading to similar distributions
at large scales. Directed polymers and associated theoretical results
thus provide useful tools to understand the statistics of optimal
paths on a complex network. It would be interesting in the future to
see if this approach can be extended to other transportation networks
or different environments, for example to the study of shortest paths
on critical percolation clusters~\cite{Barma86}, a problem with important
practical applications.

\begin{acknowledgments}
  A.P.S. thanks J.B. Deris for insightful discussions and the Gordon
  and Betty Moore Foundation for funding through a PLS
  fellowship. S.C. and M.K. acknowledge funding from the NSF, Grant
  No. DMR-12-06323.
\end{acknowledgments}


\begin{thebibliography}{99}
\bibitem{Watts98}D. J. Watts and S. H. Strogatz, Nature {\bf 393}, 440 (1998).
\bibitem{Barabasi99}A.-L. Barab\`asi and R. Albert, Science {\bf 286}, 509 (1999).
\bibitem{Albert02}R. Albert and A.-L. Barab\`asi, Reviews of Modern Physics {\bf 74}, 47 (2002).
\bibitem{Ravasz03}E. Ravasz and A.-L. Barab\`asi, Physical Review E {\bf 67}, 26112 (2003).
\bibitem{Newman03}M. E. Newman, SIAM Review {\bf 45}, 167 (2003).
\bibitem{BarYam1}J. Werfel, D.E. Ingber,  and Y. Bar-Yam, Phys. Rev. Lett. {\bf 114}, 238103 (2015).
\bibitem{BarYam2}A.J. Morales, V. Vavilala, R.M. Benito Benito, and Y. Bar-Yam, J. Royal Society Interface {\bf 14}, 20161048  (2017).
\bibitem{Barthelemy11}M. Barthélemy, Physics Reports {\bf 499}, 1 (2011).
\bibitem{Barthelemy08}M. Barthélemy and A. Flammini, Physical Review Letters {\bf 100}, 138702 (2008).
\bibitem{Cardillo06}A. Cardillo, S. Scellato, V. Latora, and S. Porta, Physical Review E {\bf 73}, 066107 (2006).
\bibitem{Lammer06}S. Lämmer, B. Gehlsen, and D. Helbing, Physica A: Statistical Mechanics and Its Applications {\bf 363}, 89 (2006).
\bibitem{Dodds00}P. S. Dodds and D. H. Rothman, Physical Review E {\bf 63}, 16115 (2000).
\bibitem{Sen03}P. Sen, S. Dasgupta, A. Chatterjee, P. Sreeram, G. Mukherjee, and S. Manna, Physical Review E {\bf 67}, 036106 (2003).
\bibitem{Hunt16}D. Hunt and V. M. Savage, Physical Review E {\bf 93}, 062305 (2016).
\bibitem{Geisberger08}R. Geisberger, P. Sanders, D. Schultes, and D. Delling, in (Springer, 2008), pp. 319–333.
\bibitem{Kardar87}M. Kardar and Y.-C. Zhang, Physical Review Letters {\bf 58}, 2087 (1987).
\bibitem{HH95}T. Halpin-Healy and Y.-C. Zhang, Physics Reports {\bf 254}, 215 (1995).
\bibitem{KPZ}M. Kardar, G. Parisi, and Y.-C. Zhang, Physical Review Letters {\bf 56}, 889 (1986).
\bibitem{foot1}{Because of this mapping, $x$ is traditionally denoted $t$ as a time direction but we stick here to the spacial
notation to avoid confusion with travel times.} 
\bibitem{FNS77}D. Forster, D.R. Nelson, and M.J. Stephen, Phys. Rev. A {\bf 16}, 732 (1977).
\bibitem{Prahofer00}M. Prähofer and H. Spohn, Physical Review Letters {\bf 84}, 4882 (2000).
\bibitem{Takeuchi11}K. A. Takeuchi, M. Sano, T. Sasamoto, and H. Spohn, Nature Scientific Reports {\bf 1} 34 (2011).
\bibitem{THH15}T. Halpin-Healy and K. A. Takeuchi, J. Stat. Phys. {\bf 160}, 794 (2015).
\bibitem{Medina89}E. Medina, T. Hwa, M. Kardar, and Y.-C. Zhang, Physical Review A {\bf 39}, 3053 (1989).
\bibitem{Peng91}C.-K. Peng, S. Havlin, M. Schwartz, and H. E. Stanley, Physical Review A {\bf 44}, R2239 (1991).
\bibitem{Kloss14}T. Kloss, L. Canet, B. Delamotte, and N. Wschebor, Physical Review E {\bf 89}, 22108 (2014).
\bibitem{Chu16}S. Chu and M. Kardar, Physical Review E {\bf 94}, 10101 (2016).
\bibitem{OSRM}D. Luxen and C. Vetter, in Proceedings of the 19th ACM SIGSPATIAL International Conference on Advances in Geographic Information Systems (ACM, New York, NY, USA, 2011), pp. 513–516.
\bibitem{Duplantier87}B. Duplantier, Physical Review B {\bf 35}, 5290 (1987).
\bibitem{Metzler02}R. Metzler, A. Hanke, P.G. Dommersnes, Y. Kantor, and M. Kardar, Phys. Rev. Lett. {\bf 88}, 188101 (2002).
\bibitem{Porto98}M. Porto, S. Havlin, H. E. Roman, and A. Bunde, Physical Review E {\bf 58}, R5205 (1998).
\bibitem{Kim91}J. Kim, M. Moore, and A. Bray, Physical Review A {\bf 44}, 2345 (1991).
\bibitem{Gueudre15}T. Gueudre, P. Le Doussal, J.-P. Bouchaud, and A. Rosso, Physical Review E {\bf 91}, 62110 (2015).
\bibitem{Pang93}N.-N. Pang and T. Halpin-Healy, Physical Review E {\bf 47}, R784 (1993).
\bibitem{Schorr03}R. Schorr and H. Rieger, The European Physical Journal B-Condensed Matter and Complex Systems {\bf 33}, 347 (2003).
\bibitem{Makse98}H. A. Makse, J. S. Andrade, M. Batty, S. Havlin, and H. E. Stanley, Physical Review E {\bf 58}, 7054 (1998).
\bibitem{Barma86}M. Barma and P. Ray, Physical Review B {\bf 34}, 3403 (1986).
\bibitem{Katsav04}E. Katzav and M. Schwartz, Physical Review E 70, 011601 (2004).
\bibitem{Song16}T. Song and H. Xia, Journal of Statistical Mechanics: Theory and Experiment 2016, 113206 (2016).
\end{thebibliography}
\end{document}